\documentstyle[aps,psfig,tighten,floats]{revtex}

\newcommand{\SS}{\scriptstyle} 

\begin{document}

\title{{\bf Finiteness of a spinfoam model for euclidean quantum
general relativity}}

\author{Alejandro Perez\thanks{Andrew Mellon Predoctoral Fellow.} \\
{\it Centre de Physique Th\'eorique - CNRS, Case 907, Luminy, F-13288
Marseille, France}; \\
{\it Physics Department, University of Pittsburgh, Pittsburgh, Pa 
15260, USA}}

\date{\today}

\maketitle

\begin{abstract}

We prove that a certain spinfoam model for euclidean quantum general
relativity, recently defined, is finite: all its all Feynman diagrams
converge.  The model is a variant of the Barrett-Crane model, and is
defined in terms of a field theory over $SO(4) \times SO(4)\times
SO(4)\times SO(4)$.

\end{abstract} 

\section{Introduction}

In reference \cite{BC}, Barrett and Crane have introduced a model for
quantum general relativity (GR).  The model is based on the
topological quantum $SO(4)$ BF theory, and is obtained by adding a
quantum implementation of the constraint that reduces classical BF
theory to euclidean GR \cite{BGGR,tqft-qg}.  To make use of the
Barrett-Crane construction in quantum gravity, two issues need to be
addressed.  First, in order to control the divergences in the sum
defining the model, the Barrett-Crane model is defined in terms of the
$q$-deformation of $SO(4)$.  In a realistic model for quantum
Euclidean GR, one would like the limit $q \rightarrow 1$ to be well
defined.

Second, the Barrett-Crane model is defined over a fixed triangulation. 
This is appropriate for a topological field theory such as BF theory,
which does not have local excitations.  But the implementation of the
BF-to-GR constraint breaks topological invariance and frees local
degrees of freedom.  The restriction of the model to a fixed
discretization of the manifold can therefore be seen only as an
approximation.  In order to capture all the degrees of freedom of
classical GR, and restore general covariance, an appropriate notion of
sum over triangulations should be implemented (see for instance
\cite{Baez}).  A novel proposal to tackle this problem is provided by
the field theory formulation of spin foam models \cite{dfkr,cm}.  In
this formulation, a full sum over arbitrary spin foams (and thus, in
particular, over arbitrary triangulations) is naturally generated as
the Feynman diagrams expansion of a scalar field theory over a group. 
The sum over spinfoams admits a compelling interpretation as a sum
over 4-geometries.  The approach represents also a powerful tool for
formal manipulations and for model building: examples of this are
Ooguri's proof of topological invariance of the amplitudes of quantum
BF theory in \cite{Ooguri} and the definition of a spinfoam model for
Lorentzian GR in \cite{acl}.

Using such framework of field theories over a group, a spinfoam model
for Euclidean quantum GR was defined in \cite{ace}.  This model
modifies the Barrett-Crane model in two respects.  First, it is not
restricted to a fixed triangulation, but it naturally includes the
full sum over arbitrary spinfoams.  Second, the natural implementation
of the BF-to-GR constraint in the field theory context fixes the
prescription for assigning amplitudes to lower dimensional simplices,
an issue not completely addressed in the original Barrett-Crane model. 
This same prescription for lower dimensional simplices amplitudes (but
in the context of a fixed triangulation) was later re-derived by Oriti
and Williams in \cite{chor}, without using the field theory.

The model introduced in \cite{ace} appeared to be naturally regulated
by those lower dimensional amplitudes.  In particular, certain
potentially divergent amplitudes were shown to be bounded in
\cite{ace}.  These results motivated the conjecture that the model
could be finite.  That is, that all Feynman diagrams might converge. 
In this letter we prove this conjecture.

This paper is not self-contained: familiarity with the formalism
defined in \cite{dfkr,cm} is assumed.  The definition of the model is
summarized in the section II; for a detailed description of the model
we refer to \cite{ace}.  In section III, a series of auxiliary results
is derived.  The proof of finiteness is given in section IV.

\section{The model}

Consider the fundamental representation of $SO(4)$, defined on
$\Re^4$, and pick a fixed direction $\hat t$ in $\Re^4$.  Let $H$ be
the $SO(3)$ subgroup of $SO(4)$ that leaves $\hat t$ invariant.  The
model is defined in terms of a field $\phi(g_1,g_2,g_3,g_4)=\phi(g_i),
i=1\ldots 4$ over $SO(4)\times SO(4) \times SO(4) \times SO(4)$,
invariant under arbitrary permutations of its arguments.  We define
the projectors $P_{h}$ and $P_g$ as
\begin{equation}  \label{ph}
P_{h}\, \phi(g_i) := \int dh_i \ {\phi}( g_ih_{i}), \ \ \ \ 
P_{g}\, \phi(g_i)
:= \int dg \ {\phi}(g_i g),
\end{equation}
where $h_i \in H$, and $g \in SO(4)$.  The model introduced in
\cite{ace} is defined by the action
\begin{equation}  \label{tope}
S[\phi]=\int dg_i \left[ P_{g} \phi(g_i) \right]^2 + \frac {1}{5!}
\int dg_i \left[ P_{g}P_{h}\phi(g_i) \right]^5,
\end{equation}
where $g_{i} \in SO(4)$, 
and the fifth power in the interaction term is notation for 
\begin{equation}
\left[\phi(g_i)\right]^5:=\phi(g_1,g_2,g_3,g_4)\ 
\phi(g_4,g_5,g_6,g_7)\
\phi(g_7,g_3,g_8,g_9)\ \phi(g_9,g_6,g_2,g_{10}) \ 
\phi(g_{10},g_8,g_5,g_1).
\end{equation}

$P_g$ projects the field into the space of gauge invariant fields,
namely, those such that $\phi(g_1,g_2,g_3,g_4)
=\phi(g_1g,g_2g,g_3g,g_4g)$ for all $g \in SO(4)$.  The projector
$P_{h}$ projects the field over the linear subspace of the fields that
are constants on the orbits of $H$ in $SO(4)$.  When expanding the
field in modes, that is, on the irreducible representations of
$SO(4)$, this projection is equivalent to restricting the expansion to
the representations in which there is a vector invariant under the
subgroup $H$ (because the projection projects on such invariant
vectors).  The representations in which such invariant vectors exist
are the ``simple", or ``balanced", representations namely the ones in
which the spin of the self dual sector is equal to the spin of the
antiselfdual sector%
\footnote{ Representations of $SO(4)$ can be labeled by two integers
$(n_1,n_2)$.  In terms of those integers, the dimension of the
representation is given by $\Delta_{(n_1,n_2)}=(n_1+1)(n_2+1)$. 
Simple representations are those for which $n_1=n_2=N$.}.
In turn, the simple representations are the ones whose generators have
equal selfdual and antiself dual components, and this equality, under
identification of the $SO(4)$ generator with the $B$ field of $BF$
theory is precisely the constraint that reduces $BF$ theory to GR.
Alternatively, this constraint allows one to identify the generators
as bivectors defining elementary surfaces in 4d, and thus to interpret
the coloring of a two-simplex as the choice of a (discretized) 4d
geometry \cite{Reisenberger,Barbieri,BC,Baez}.  

Using the Peter-Weyl theorem one can write the partition function of
the theory
\begin{equation}  \label{fe}
Z:=\int {\cal D}\phi\ e^{-S[\phi]}
\end{equation}
as a perturbative sum over the amplitudes $A(J)$ of Feynman diagrams
$J$.  This computation is performed in great detail in \cite{ace},
yielding 
\begin{equation} 
Z = \sum_{J}\ A(J) = \sum_{J}\ \sum_{N}\ \prod_{f\in J} 
\Delta_{N_{f}} \ \prod_{e\in J}
A_e(N_e) \ \prod_{v\in J} A_v(N_v).  \label{Z}
\end{equation}
The first summation is over pentavalent 2-complexes $J$, defined
combinatorially as a set of faces $f$, edges $e$ and vertices $v$, and
their boundary relations%
\footnote{The Feynman diagrams of the theory are obtained by
connecting the five-valent vertices with propagators.  At the open
ends of propagators and vertices there are the four group variables
corresponding to the arguments of the field.  A 2-complex is given by
a certain vertices-propagators topology plus a fixed choice of a
permutation on each propagator (see \cite{ace}).  Strictly speaking,
Feynman diagrams of the field theory are given by the five valent
graphs.  On this 1-skeleton, a 2-complex is defined by each one of the
permutations.  However, following \cite{ace}, we will call here
``Feynman diagram" each one of such 2-complexes.}.
The second sum is over simple $SO(4)$ representations $N$ coloring the
faces of $J$.  $\Delta_N$ is the dimension of the simple
representation $N$.  The amplitude $A_e(N)$ is a function of the four
colors that label the corresponding faces bounded by the edge.  It is
explicitly given by
\begin{equation}  \label{xii}
A_e=\frac {\Delta_{N_1, \ldots,
N_{4}}}{\left(\Delta_{N_1}\dots\Delta_{N_4}\right)^{2}},
\end{equation}
where ${\Delta_{N_1, \ldots, N_{4}}}$ is the dimension of the space of
the intertwiners between the four representations $N_1, \ldots,
N_{4}$\cite{ace}.  The vertex amplitude $A_v$ is the Barrett-Crane
vertex amplitude, which is a function of the ten colors of the faces
adjacent to the 5-valent vertex of $J$.  The Barrett-Crane vertex
amplitude can be written as a combination of $15j$ symbols.  However,
as it was shown by Barrett in \cite{intbc}, it can also be express as
an integral over five copies of the 3-sphere -- a representation with
a nice geometrical interpretation.  This representation is better
suited for our purposes so we give it here explicitly 
\begin{eqnarray}  \label{vv}
&& A_v(N_{\scriptscriptstyle 1}, \dots N_{{10}}) = \int_{(S^3)^5} 
dy_1 \dots
dy_5 \ \ K_{N_1}(y_1,y_5)\ K_{N_2}(y_1,y_4)\ K_{N_3} 
(y_1,y_3)\ K_{N_4}(y_1,y_2) 
\nonumber \\
&& \ \ \ \ \ \ \ \ \ \ \ \ \ \ \ \ \ \ \ \ K_{N_5}(y_2,y_5)
\ K_{N_6}(y_2,y_4)\ K_{N_7}(y_2,y_3)\  K_{N_8}(y_3,y_5)\ 
K_{N_9}(y_3,y_4)\  K_{N_{10}}(y_4,y_5),
\end{eqnarray}
where $dy$ denotes the invariant normalized measure on the sphere.  If
we represent the points in the 3-sphere as unit-norm vectors $y^{\nu}$
in $\Re^4$, and we define the angle $\Theta_{ij}$ by ${\rm
cos}(\Theta_{ij})=y^{\mu}_iy^{\nu}_i \delta_{\mu \nu}$, then the
kernel $K_N $ is given by
\begin{equation}  \label{kr}
K_N(y_i,y_j)=\frac{{\rm sin}\left((N+1)\Theta_{ij}\right)}{{\rm sin}%
(\Theta_{ij})}. 
\end{equation}
This is a smooth bounded functions on $S^3 \times S^3$, with maximum
value $N+1$.

\section{Bounds} 

Our task is to prove convergence of the Feynman integrals of the
theory.  In the mode expansion, potential divergences appear in the 
sum over representations $N$.  Therefore we need to analyze the
behavior of vertex and edge amplitudes for large values of $N$.

An arbitrary point $y\in S^{3}$ can be written in spherical 
coordinates as 
\begin{equation}
y=\left( {\rm cos}(\psi ),{\rm sin}(\psi ){\rm sin}(\theta ){\rm
cos}(\phi ), {\rm sin}(\psi ){\rm sin}(\theta ){\rm sin}(\phi ),{\rm
sin}(\psi ){\rm cos} (\theta )\right) ,
\end{equation}
where $0 \le \psi \le \pi $, $0 \le \theta \le \pi $, and $0\le \phi
\le 2\pi $. The invariant normalized measure in this coordinates 
is 
\begin{equation}
dy=\frac{1}{2\pi^2 }\ {\rm sin}^2(\psi )\, {\rm sin}(\theta )\ d\psi\, d\theta 
\, d\phi .
\label{mea}
\end{equation}
Using the gauge invariance of the vertex, the invariance of the three
sphere under the action of $SO(4)$ and the normalization of the
invariant measure, we can compute (\ref{vv}) by dropping one of the
integrals and fixing one point on $S^{3}$, say $y_{1}=(1,0,0,0)$. 
Thus equation (\ref{vv}) becomes
\begin{eqnarray}\label{sisi}
&& A_v(N_{\scriptscriptstyle 1}, \dots N_{{10}}) =\frac{1}{16} \int 
d\psi_2 \dots d\psi_5
\ d\Omega_2 \dots d\Omega_5 \ 
{\rm sin}\left({\SS (N_1+1)}\psi_5\right) 
{\rm sin}\left({\SS (N_2+1)}\psi_4\right)
{\rm sin}\left({\SS (N_3+1)}\psi_3\right)
{\rm sin}\left({\SS (N_4+1)}\psi_2\right)
\nonumber \\
&& \ \ \ \ \ \ \ \ \ \ \ \ \ \ \ \ \ \ \ \ K_{N_5}(y_2,y_5)
K_{N_6}(y_2,y_4)K_{N_7}(y_2,y_3) K_{N_8}(y_3,y_5)K_{N_9}(y_3,y_4)
K_{N_{10}}(y_4,y_5),
\end{eqnarray}
where $d\Omega_i$ is the normalized measure on the 2-sphere 
$\psi_i=\;$constant.
Now we bound the Barrett-Crane amplitude using that $K_N \le N+1= 
\sqrt \Delta_N$,
namely
\begin{eqnarray}\label{depe}
\nonumber 
\left|A_v(N_{\scriptscriptstyle 1}, \dots N_{{10}})\right| && \le 
\left(\frac{2}{\pi}\right)^4 
\sqrt{\Delta_{N_5}\Delta_{N_6}\Delta_{N_7}\Delta_{N_8}
\Delta_{N_9}\Delta_{N_{10}}} 
\ \ 
\prod^{4}_{i=1} \int^{\pi}_0 d\psi_i\ \left| {\rm sin}\left({\SS 
(N_i+1)}\psi_i\right){\rm sin}(\psi_i) \right|  
 \\
&&
\le \ \ \ \; 
\sqrt{\Delta_{N_5}\Delta_{N_6}\Delta_{N_7}\Delta_{N_8}\Delta_{N_9}
\Delta_{N_{10}}}.
\end{eqnarray}
The argument is obviously independent of the choice of the six colors
in (\ref{depe}).  Weaker versions of the bound in which more colors
are included also hold.  In particular if we directly bound the
$K_N$'s in (\ref{vv}) we obtain that the absolute value of the
amplitude is bounded by the square root of the product of the ten
dimensions.  More in general, let $I_{10}(k)$, with $k$ taking the
values $0,1,2,3,4$, be an arbitrary subset of $\{1,2,\dots 10 \}$ with
$10-k$ elements.  Then the following bound holds for any $I_{10}(k)$
\begin{eqnarray}\label{cota}
\left|A_v(N_{\scriptscriptstyle 1}, \dots N_{{10}})\right|\le
\prod_{i \in {I_{10}(k)}} \sqrt{\Delta_{N_i}}\ \ \ \ \ \forall \ 
I_{10}(k),\ \ k=0,1,2,3.
\end{eqnarray}
For $k=4$ we recover (\ref{depe}).

In \cite{williams}, Barrett and Williams have analyzed the asymptotic
behavior of the oscillatory part of the amplitude, in connection to
the classical limit of the theory.  We add here some information on
the asymptotic behavior of the magnitude of (\ref{vv}).  The results
of this paragraph will not be used in the rest of the paper; we
present them since they follow naturally from our previous
considerations.  Equation (\ref{sisi}) can be rewritten as
\begin{equation}
A_{v}(N_{\scriptscriptstyle 1},\dots N_{{10}})=\int {\rm 
sin}((N_{1}+1)\psi
_{5})F(N_{\scriptscriptstyle 2},\dots N_{{10}};\psi _{5})d\psi _{5},
\end{equation}
where $F(N_{\scriptscriptstyle 2},\dots N_{{10}};\psi _{5})$ is a
smooth bounded function in $[0,\pi ]\in \Re $, since $F$ is given by
an integral of smooth and bounded functions on a compact space. 
Therefore, the following limit holds
\begin{equation}
\lim \limits_{N_1 \rightarrow \infty} A_{v}(N_{\scriptscriptstyle 
1},\dots N_{{10}})=0.
\end{equation} 
The same argument can be used to prove the following stronger result
\begin{equation}
\lim \limits_{N_1,\dots N_4 \rightarrow \infty} 
A_{v}(N_{\scriptscriptstyle 1},\dots N_{{10}})=0,
\end{equation} 
where, in the limit, the four colors are taken simultaneously to
infinity.  Clearly, these equations are valid for any choice of the
$N_i$'s.

Finally, it is easy to verify the following inequalities for the
dimension of the space of intertwiners between four
representations converging at an edge (see \cite{ace})
\begin{equation}
{\Delta_{N_{1}, \ldots,N_{{4}}}} \le \sqrt{\Delta_{N_{i}}},
\end{equation}
which holds for any $i=1,2,3,4$.  Using this expression, we can
construct weaker bounds in the spirit of (\ref{cota}).  We define the
set $I_{4}(k)$ ($k=0,1,2,3$ ) as an arbitrary subset of $\{1,2,3,4\}$
with $4-k$ elements.  Then the following inequalities hold for any
$I_4(k)$
\begin{equation}\label{dd}
{\Delta_{N_{1}, \ldots,N_{{4}}}} \le \prod_{I_4(k)} 
\sqrt{\Delta_{N_i}} \ \ \ \ \ 
\forall \ {I}_{4}(k), \ \ k=0,1,2,3. 
\end{equation}

\section{Proof of finiteness}

We have now all the tools for proving the finiteness of the model.  We
will construct a finite bound for the amplitudes $A(J)$, defined in
(\ref{Z}), of arbitrary Feynman diagrams of the model.  Inserting the
inequalities (\ref{dd}) and (\ref{cota}), both with $k=0$, in the
definition of $A(J)$, we obtain a bound for the amplitude of an
arbitrary pentavalent 2-complex $J$. Namely,
\begin{equation}\label{lala}
|A(J)| \le \prod_{f \in J}  \sum_{N_f}\left( \Delta_{N_{f}} 
\right)^{1-2 n_f+n_f/2+n_f/2} = 
\prod_{f \in J}  \sum_{N_f}\left( \Delta_{N_{f}} 
\right)^{1-n_f},
\end{equation}
where $n_f$ denotes the number of edges of the face $f$.
The term $\Delta^{1-n_f}_N$ in (\ref{lala}) comes from
various contributions.  First, we have $\Delta_N$ from the face
amplitude.  Second, we have $\Delta^{-2 n_f}_N$ and $\Delta^{n_f/2}_N$
from the denominators of the $n_f$ edge amplitudes (\ref{xii}) and the bounds for
the corresponding numerators (\ref{dd}) respectively.  Finally, we
have $\Delta^{n_f/2}_N$ from the bounds for the $n_f$ vertex
amplitudes (\ref{cota}).  

If the 2-complex contains only faces with more than one edge, then the
previous bound for the amplitude is finite.  More precisely, if all
the $n_f$'s are such that $n_f \ge 2$ then $1- n_f \le -1$, and using
the fact that $\Delta_N=(N+1)^2$ the sum on the RHS of the last
equation converges.  

\begin{figure}[tbp]
\centerline{{\psfig{figure=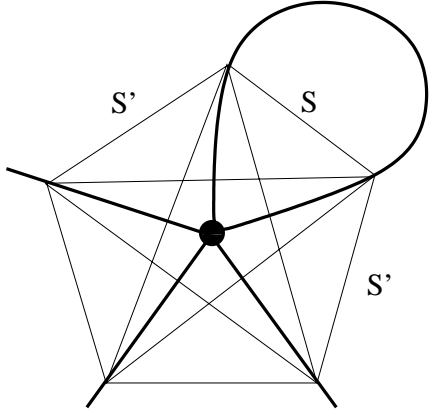,height=6cm}}} \bigskip
\caption{The ``singular'' face}
\label{loop}
\end{figure}

On the other hand, if some of the $n_f$ are equal to 1, then the right
hand side of (\ref{lala}) diverges, and therefore for this case we need
a stricter bound, involving stronger inequalities.  This can be done
as follows.  Notice that every time a 2-complex contains a face whose
boundary is given by a single edge, the same edge must be part of the
boundary of another face, bounded by more than one edge.  In Fig.\, 1
an elementary vertex of a 2-complex containing such a face is shown. 
The thick lines represent edges converging at the vertex, each of them
is part of the boundary of four faces.  To visualize those faces we
have drawn in thin lines the intersection of the vertex diagram with a
3-sphere.  There is a face bounded by a single edge.  Its intersection
with the sphere is denoted by $S$.  Notice that the surface
intersecting the sphere in $S^{\prime}$ will have a boundary defined
by at least two edges.  Notice also that a single vertex can have a
maximum of two such peculiar faces. 

Therefore, using (\ref{dd}) for $k=1$ we can choose to bound the
numerator in (\ref{xii}) with the colors corresponding to the three
adjacent faces like $S^{\prime}$ in Fig.\, 1.  If $N_1$ denotes the
color of the face $S$ then we construct the bound for the amplitude
using
\begin{equation}
{\Delta_{N_1, \ldots,N_{4}}} \le 
\sqrt{\Delta_{N_2}\Delta_{N_3}\Delta_{N_4}}.
\end{equation}
One of the square roots would be sufficient to bound the edge
amplitude, but the symmetry in the previous expression simplifies the
construction of the bound for the amplitude of an arbitrary 2-complex. 
Then use (\ref{cota}) for $k=1$ (or $k=2$) to bound the vertex
amplitude corresponding to a vertex containing one (respectively two)
face(s) whose boundary is given by only one edge.  In this way, we can
exclude the color corresponding to these ``singular'' faces from the
bounds corresponding to the vertex and the denominator of the edge
amplitude.  Thus these faces contribute to the bound as $\Delta_N$
(face amplitude) times $\Delta^{-2}_N$ (from the denominator of the
single edge amplitude), i.e., as $\Delta_{N}^{-1}$.  We keep using 
(\ref{dd}), and (\ref{cota}) for $k=0$ for faces with $n_f>1$ and
vertices containing no faces with $n_f=1$.  If we denote by $\{
f_{(n_f > 1)}\}$ and $\{ f_{(n_f=1)}\}$ the set of faces with more
than one edge and one edge respectively, the general bound is finally
given by
\begin{equation}\label{oui}
|A(J)| \le \prod_{f_{(n_f > 1)} \in J}\ \ \sum_{N_f} \left( 
\Delta_{N_{f}} 
\right)^{1- n_f} \ \ \prod_{f_{(n_f = 1)} \in J}\ \ \sum_{N_f} \left( 
\Delta_{N_{f}} 
\right)^{-1} \le \prod_{f \in J}\ \ \sum_{N_f} \left( \Delta_{N_{f}} 
\right)^{-1} = \left(\zeta(2)-1\right)^{F_J}\approx (0.6)^{F_J},
\end{equation}
where $F_J$ denotes the number of faces in the 2-complex $J$, and
$\zeta$ denotes the Riemann zeta function.  This concludes the proof
of the finiteness of the amplitude for any 2-complex $J$ $\Box$

\section{Discussion}

Equation (\ref{oui}) proves that there are no divergent amplitudes in
the field theory defined by (\ref{tope}).  This field theory was
defined in \cite{acl} as a model for euclidean quantum gravity based
on the implementations of the constraints that reduce $SO(4)$ BF
theory to euclidean GR. The corresponding BF topological theory is
divergent after quantization: regularization is done ad-hoc by
introducing a cut-off in the colors or, more elegantly, by means of
the quantum deformation of the group ($SO(4) \rightarrow SO_q(4)$). 
Remarkably, the implementation of the constrains that give the theory
the status of a quantum gravity model automatically regularizes the
amplitudes.

Another encouraging result comes from the fact that  according to 
equation (\ref{oui}) contributions of Feynman diagrams decay 
exponentially with the number of faces. This might be useful for 
studying the convergence of the full sum over 2-complexes. 

\section{Acknowledgments}  

To Carlo Rovelli.

\end{document}